\documentclass{webofc}

\usepackage[varg]{txfonts}   
\usepackage{hyperref}
\usepackage{url}
\usepackage{lineno}  
\hypersetup{colorlinks=true,citecolor=blue,urlcolor=blue,linkcolor=blue}
\begin{document}
%
%
\title{Studies of jet quenching in O+O collisions at $\sqrt{s_{\rm NN}}$ = 200 GeV by STAR}
%
%

\author{\firstname{Sijie} \lastname{Zhang}\inst{1}\fnsep\thanks{\email{sijiezhang@mail.sdu.edu.cn}} 
(for the STAR Collaboration)
}

\institute{Institute of Frontier and Interdisciplinary Science, Shandong University,
Qingdao, Shandong 266237, China
          }

\abstract{
Jet quenching is a well-established probe of the QGP in large collision systems such as Au+Au and Pb+Pb, but is not observed in smaller p+A collisions despite the presence of collectivity. This makes it important to study its dependence on system size. O+O collisions offer an ideal opportunity, bridging the gap between small and large systems.
In these proceedings, we present the measurements of the following observables related to jet quenching: inclusive charged hadron $R_{\rm AA}$ and $R_{\rm CP}$ at high transverse momentum ($p_{\rm T}$), inclusive jet $R_{\rm CP}$, and semi-inclusive hadron+jet $I_{\rm CP}$, using O+O collisions data collected with the STAR detector in 2021 at $\sqrt{s_{\rm NN}} = 200$ GeV. 
For jet measurements, combinatorial background is subtracted based on the event mixing technique, while background fluctuations and detector effects are corrected via unfolding. 
Inclusive charged hadron and jet $R_{\rm CP}$ and h+jet $I_{\rm CP}$ are found to be below unity, indicating suppression relative to peripheral events.
}
\maketitle

\section{Introduction}
\label{intro}

Jet quenching, the modifications to the energy and substructure of high-energy parton showers in the Quark Gluon Plasma (QGP), serves as a key experimental tool to probe properties of QGP in relativistic heavy-ion collisions. The quenching effect has been observed with single hadrons, inclusive and semi-inclusive h+jet in Au+Au collisions~\cite{STAR:2017hhs,STAR:2020xiv}.

For small systems, while collectivity is observed in $p$+A and $d$+A collisions~\cite{STAR:2015kak,STAR:2014qsy,STAR:2023wmd}, clear evidence of jet quenching remains elusive. In $p$+Au collisions, the STAR collaboration observed h+jet suppression in high event activity collisions compared to low event activity collisions. However, this suppression was observed on both the trigger-side and recoil-side, and the dijet momentum imbalance distribution shows no significant modification in high event activity (EA) events. This suggests the suppression is not due to jet quenching but rather due to anti-correlation between EA and momentum-transfer ($Q^2$)~\cite{STAR:2024nwm}.

The O+O collision system presents an opportunity to bridge the gap between small and large systems and investigate the system size dependence of jet quenching. The participant nucleon number ($N_{\rm part}$) distribution of O+O collisions overlaps with $p$+Au and $d$+Au systems but has a flatter distribution, suggesting less centrality fluctuations. Long-range ridge correlations have been observed in central O+O collisions, providing clear evidence of collective flow~\cite{Yan2025QuarkMatter}. Theoretical calculations using nuclear parton distribution functions (nPDF) have been performed for O+O collisions, providing baseline expectations without quenching effects~\cite{Gebhard:2024flv}.

The STAR detector completed data collection for O+O collisions at $\sqrt{s_{\rm NN}} = 200$ GeV in 2021. The Time Projection Chamber (TPC) provides tracking information for charged particles with pseudorapidity $|\eta| < 1.5$ and full azimuthal coverage, enabling charged-particle jet reconstruction. The Event Plane Detector (EPD) covers the forward pseudorapidity range $\eta$ = 2.1—5.1 and is used for centrality definition.

\section{Analysis Procedure}
\label{sec:analysis}

This analysis investigates several complementary jet quenching observables. First, high-$p_{\rm T}$ charged particles serve as jet proxies, with $R_{\rm AA}$ and $R_{\rm CP}$ defined in Eq.~\ref{eq:RAA,RCP,Icp}, where $Y_{\rm AA}$, $Y_{\rm pp}$, $Y_{\rm central}$, and $Y_{\rm peripheral}$ denote the yields in A+A, $p$+$p$, central, and peripheral events, respectively. Second, jet reconstruction better represents initial partons and enables inclusive jet yield measurements, and the inclusive jet $R_{\rm CP}$ is also reported.

Both observables depend on the number of nucleon-nucleon collisions ($N_{\rm coll}$) from the Glauber model. Alternatively, the semi-inclusive h+jet method selects high-$p_{\rm T}$ charged hadrons as triggers ($7 < p_{\rm T}^{\rm trig} < 30$~GeV/$c$) and measures recoil jet yields with $|\phi_{\rm trig} - \phi_{\rm jet} - \pi| < \pi/4$. Since recoil jet yields are normalized by the number of trigger hadrons, the $I_{\rm CP}$ measurement avoids reliance on the model-dependent $N_{\rm coll}$ parameter (~Eq.~\ref{eq:RAA,RCP,Icp}).

\begin{equation}
     \label{eq:RAA,RCP,Icp}
     R_{\rm AA}=\frac{Y_{\rm AA}}{\langle N_{\rm coll} \rangle Y_{\rm pp}}, \quad
     R_{\rm CP}=\frac{\langle N_{\rm coll} \rangle^{\rm peripheral}Y_{\rm central}}{\langle N_{\rm coll} \rangle^{\rm central}Y_{\rm peripheral}}, \quad
     I_{\rm CP}=\frac{Y_{\rm central}}{Y_{\rm peripheral}}
\end{equation}

Jets are reconstructed using the anti-k$_{\rm T}$ algorithm and their transverse momentum is corrected for the average background  density ($\rho$) as $p_{\rm T,\,\text{jet}}^{\text{reco,ch}} = p_{\rm T,\,\text{jet}}^{\text{raw,ch}} - \rho \cdot A$, where A is the jet area.
Mixed events are constructed by randomly selecting tracks from different real events, allowing for combinatorial background estimation.
Moreover, the $p_{\rm T,\,\text{jet}}^{\text{reco,ch}}$ spectra include detector effects and background fluctuations, which are corrected through iterative Bayesian unfolding.

\section{Results}
\label{sec:results}

\begin{figure}[h]
     \vspace{-8pt}
     \centering
     \sidecaption
     \includegraphics[width=0.4\textwidth, trim=-10 0 -60 0, clip]{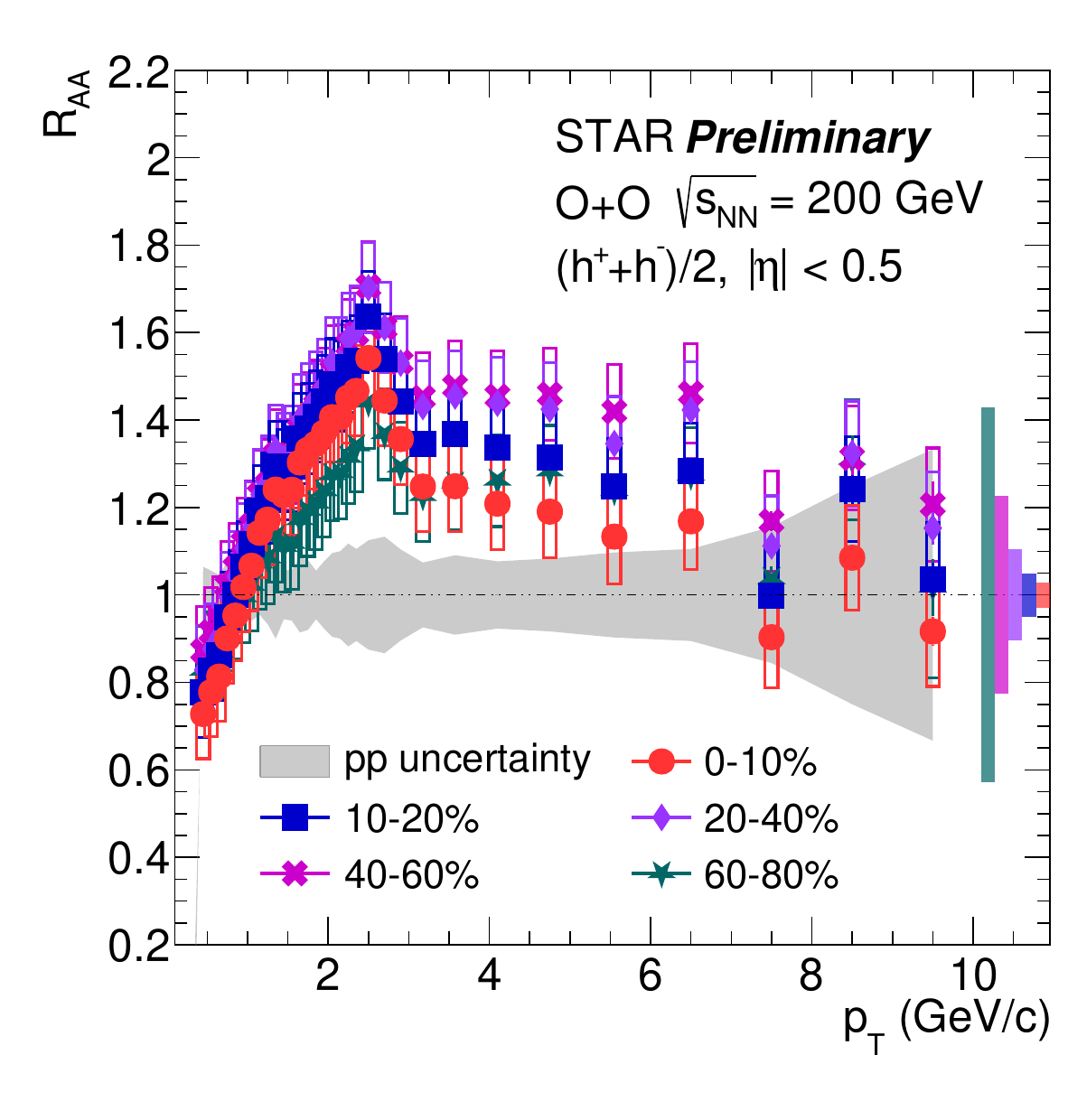}
     \vspace{-8pt}
     \caption{Inclusive charged hadron $R_{\rm AA}$ as a function of $p_{\rm T}$ in different centrality classes of O+O collisions at $\sqrt{s_{\rm NN}}$ = 200 GeV. Vertical bars and boxes around data points display statistical and systematic uncertainties, respectively.}
     \vspace{-12pt}
     \label{fig-RAA}       
\end{figure}

The inclusive charged hadron $R_{\rm AA}$ measurements, using $p$+$p$ collisions as the reference, are shown in Fig.~\ref{fig-RAA}. The uncertainty of $p$+$p$ measurement and $N_{\rm coll}$ are represented by gray shading and colored boxes, respectively. 
In the region where $p_{\rm T} < 2.5$ GeV/$c$, $R_{\rm AA}$ increases with $p_{\rm T}$ and could be explained by cold nuclear matter effects~\cite{STAR:2003pjh}. For $p_{\rm T} > 2.5$ GeV/$c$, the central values of $R_{\rm AA}$ decrease slightly. At high $p_{\rm T} > 7$ GeV/$c$ in 0--10\% collisions, $R_{\rm AA}$ is approximately unity with about 15\% uncertainty. 

\begin{figure}[h]
     \vspace{-5pt}
     \centering
     \sidecaption
     \includegraphics[width=0.3\textwidth, trim=10 15 0 0, clip]{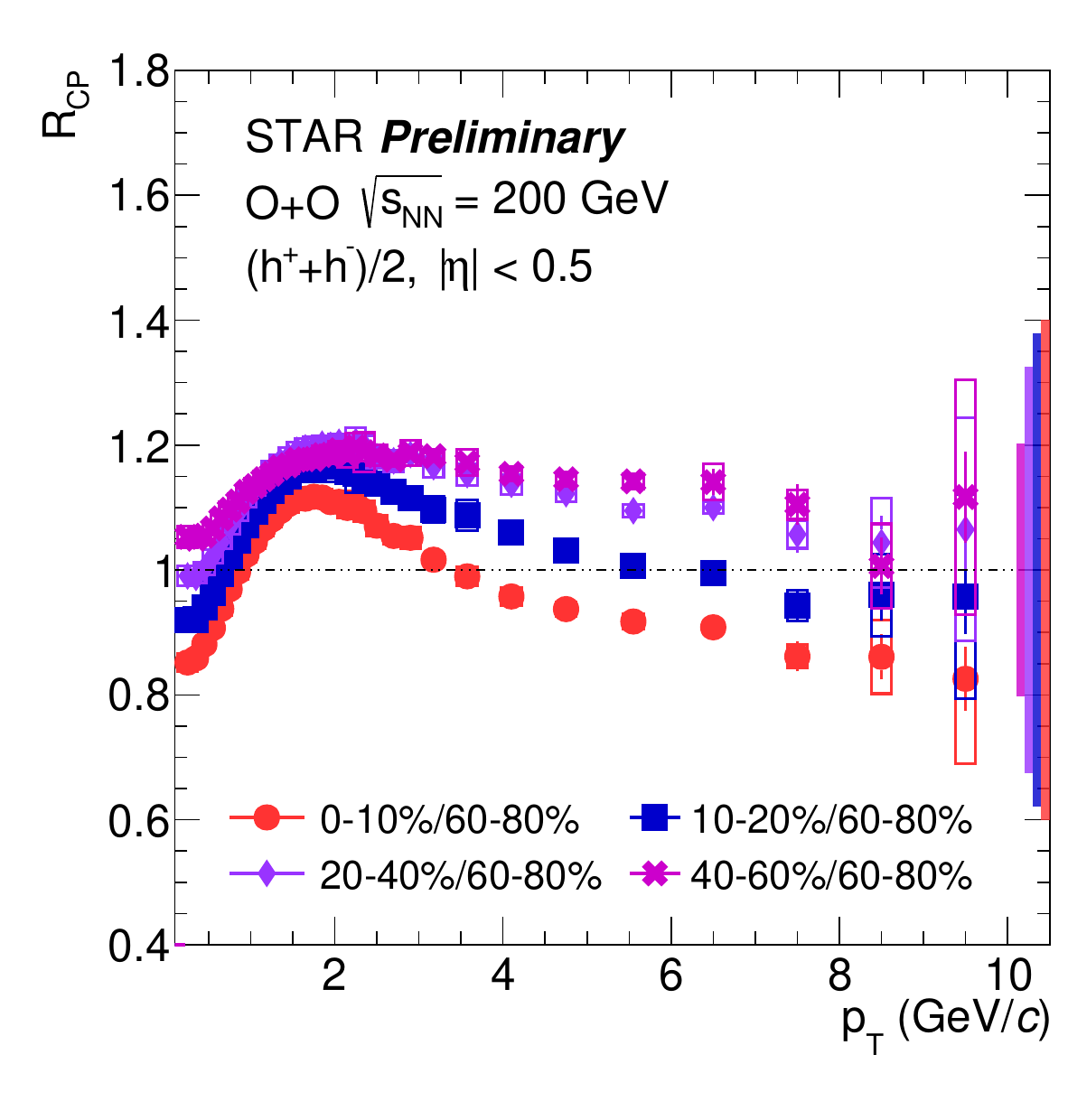}
     \includegraphics[width=0.3\textwidth, trim=50 45 50 45, clip]{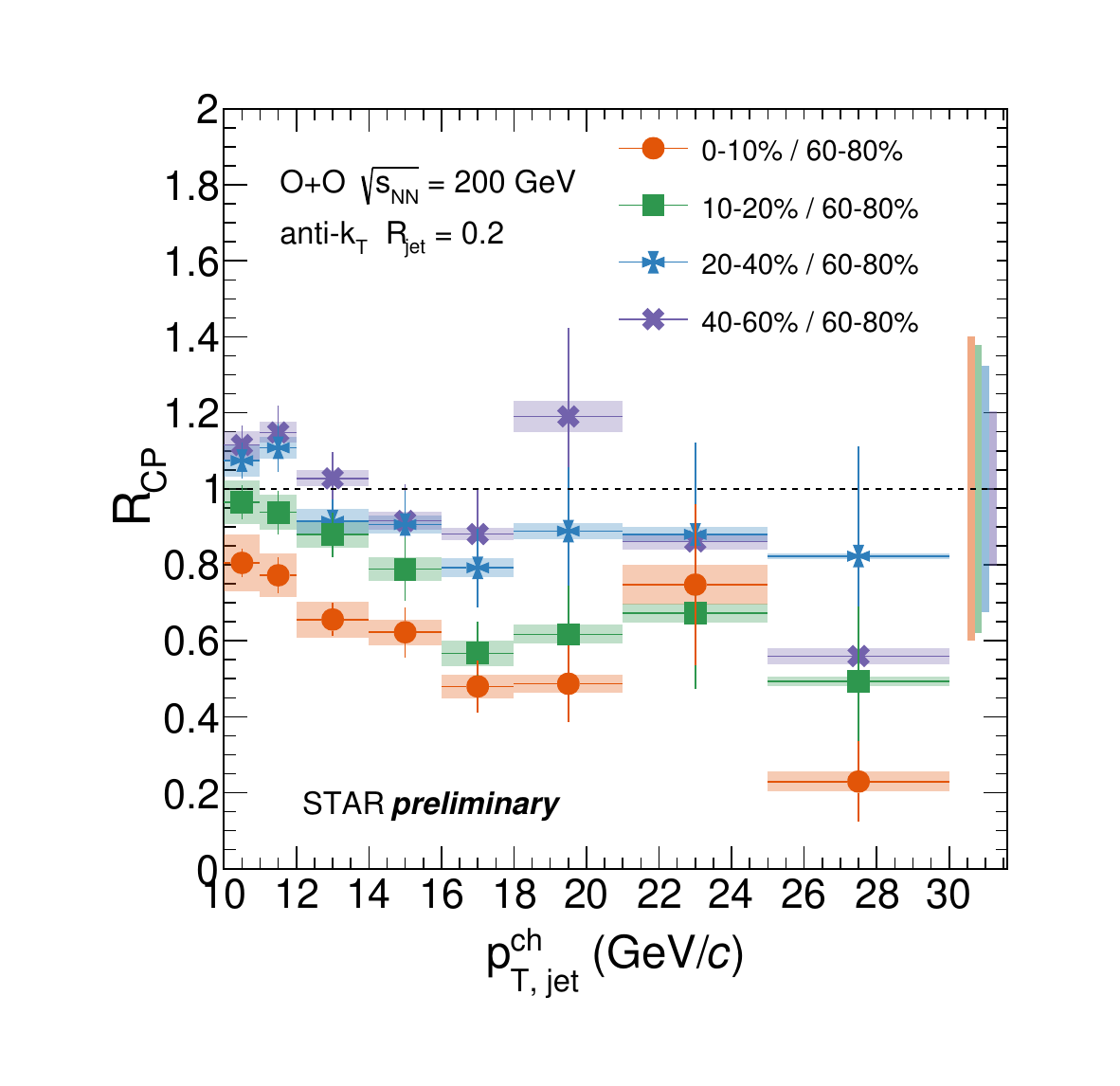}
     \vspace{-5pt}
     \caption{Inclusive charged hadron (left) and jet (right) $R_{\mathrm{CP}}$ as a function of $p_{\mathrm{T}}$ in different centrality classes of O+O collisions at $\sqrt{s_{\mathrm{NN}}}$ = 200 GeV. Vertical bars and boxes around data points display statistical and systematic uncertainties, respectively.}
     \vspace{-10pt}
     \label{fig-Rcp}       
\end{figure}

Jet measurements extend the analysis to higher $p_{\rm T}$ regions and provide better access to parton kinematics. 
Figure~\ref{fig-Rcp} left and right panels shows the $R_{\rm CP}$ of hadrons and jets, respectively. Although fragmentation means that hadrons and jets are not directly comparable, both exhibit a suppression trend at high $p_{\rm T}$.
Overall, $R_{\rm CP} < 1$ at high $p_{\rm T}$ for the 0--10\% centrality, but large uncertainties in $N_{\rm coll}$ calculations---shown as a colored box at the rightmost part of the figure---make it difficult to draw definitive conclusions about jet quenching.

\begin{figure}[h]
     \centering
     \sidecaption
     \includegraphics[width=0.3\textwidth, trim=50 45 50 45, clip]{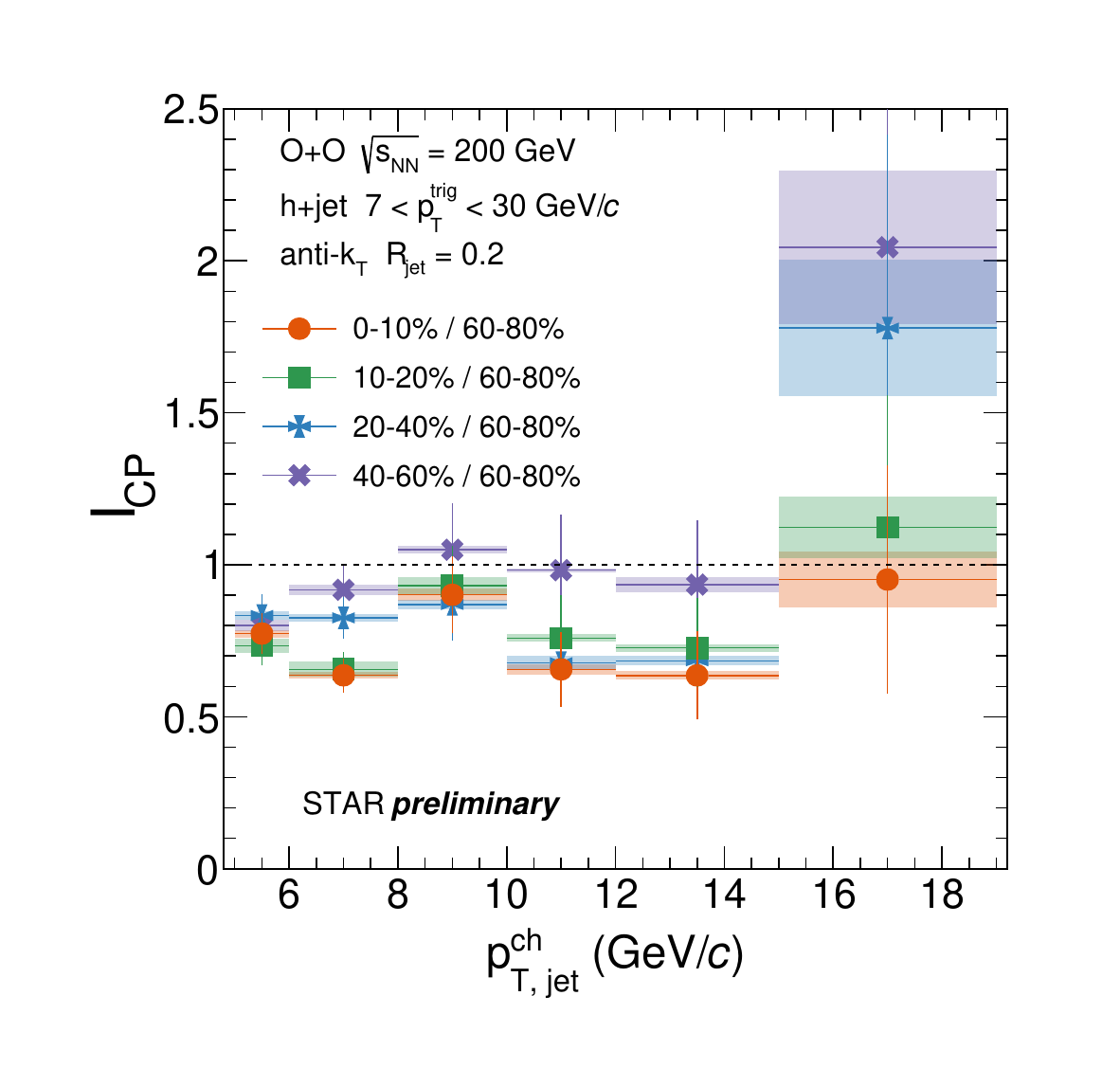}
     \includegraphics[width=0.3\textwidth, trim=50 45 50 45, clip]{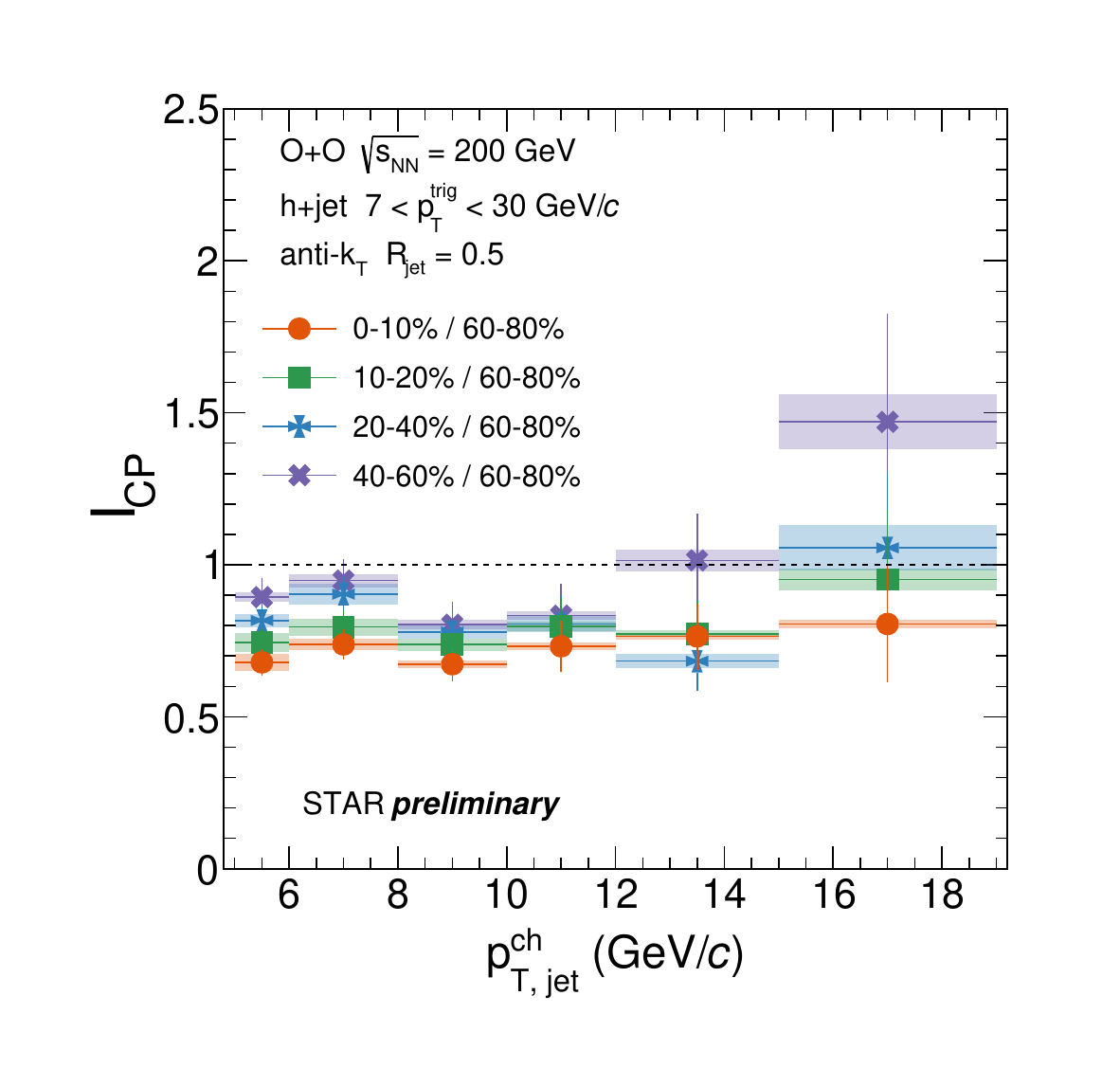}
     \caption{Recoil jet $I_{\mathrm{CP}}$ as a function of $p_{\mathrm{T,jet}}^{\mathrm{ch}}$ for two jet radii of $R=0.2$ (left) and 0.5 (right) in different centrality classes of O+O collisions at $\sqrt{s_{\mathrm {NN}}}$ = 200 GeV. Vertical bars and boxes around data points display statistical and systematic uncertainties, respectively.}
     \vspace{-8pt}
     \label{fig:hjet_Icp_ref_60_80}
 \end{figure}

The semi-inclusive h+jet $I_{\rm CP}$ measurements provide important cross-checks that avoid model-dependent uncertainties. Results are presented for both $R = 0.2$ and $R = 0.5$ jet resolution parameters in Fig.~\ref{fig:hjet_Icp_ref_60_80}, with different colors and markers representing different centrality bins. The measurements show no significant difference between $R = 0.2$ and $R = 0.5$ results, and the $I_{\rm CP}$ values are consistently < 1. Since these measurements are not affected by $N_{\rm coll}$ uncertainties, the observed suppression suggests yield suppression in O+O collisions.

\begin{figure}[h]
     \sidecaption
     \vspace{-8pt}
     \centering
     \includegraphics[width=0.48\textwidth, trim=0 5 0 30, clip]{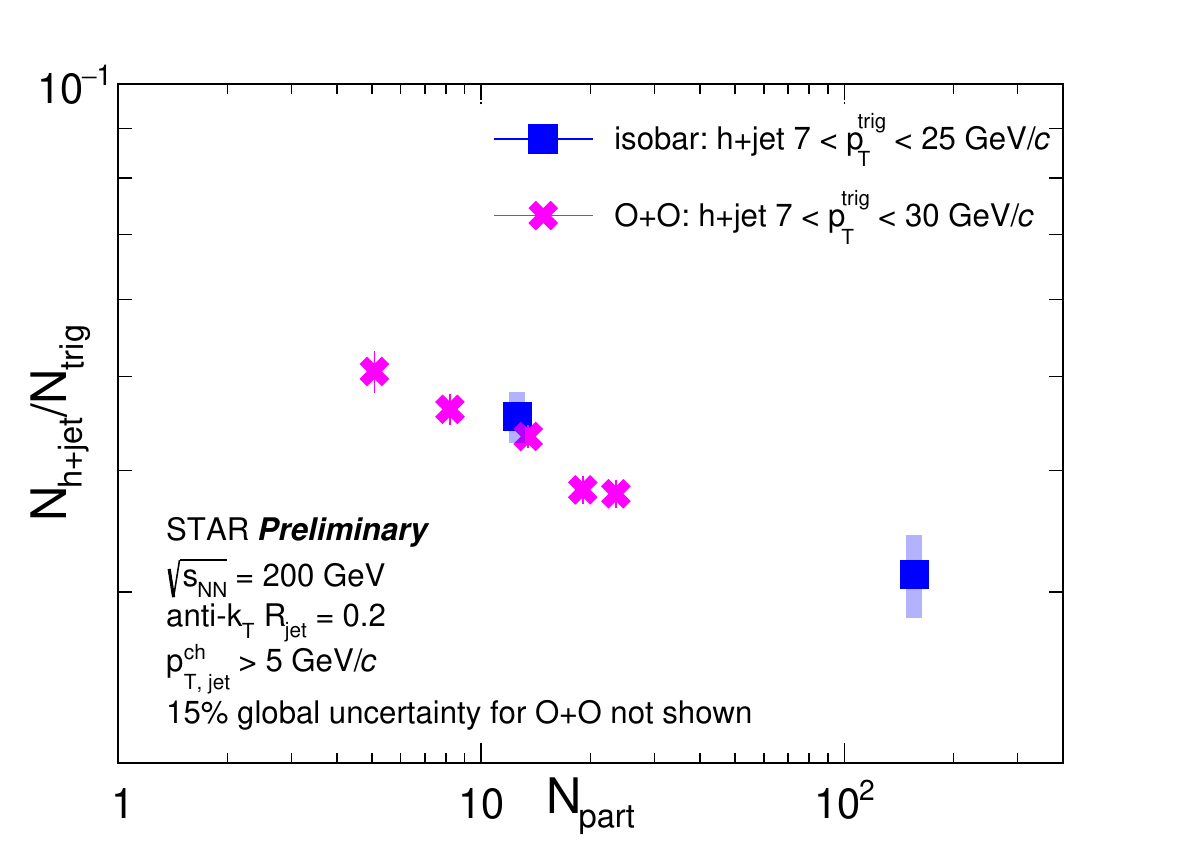}
     \vspace{-8pt}
     \caption{Recoil jet yields of O+O and Ru+Ru / Zr+Zr isobar collisions at $\sqrt{s_{\mathrm {NN}}}$ = 200 GeV, integrated for $p_{\mathrm{T,jet}}^{\mathrm{ch}} > 5$ GeV/$c$, as a function of the number of participating nucleons ($N_{\text{part}}$), from peripheral to central (left to right).}
     \vspace{-8pt}
     \label{fig-Npart}       
\end{figure}

A comparison between O+O and isobar collisions, illustrated in Fig.~\ref{fig-Npart}, shows the recoil jet yields as a function of $N_{\rm part}$, indicating that the yields decrease with increasing $N_{\rm part}$.
Since the yield of peripheral events differs between O+O and isobar systems, even if they have the similar $I_{\rm CP}$, their central events may still have experienced different levels of quenching.

The observation of $I_{\rm CP} < 1$ in O+O collisions raises important questions about the interpretation of yield suppression. It remains necessary to investigate whether this suppression indicates jet quenching caused by QGP formation or arises from other contributing factors. 

\color{black}

\section{Summary}
\label{sec:summary}

We reported measurements of several observables that indicate jet yield suppression in central relative to peripheral O+O collisions at $\sqrt{s_{\rm NN}}$ = 200 GeV.
The charged hadron $R_{\rm AA}$ is close to unity with about 15\% uncertainty. 
Charged hadron and jet $R_{\rm CP}$ values are below unity but are subject to large uncertainties, primarily due to $N_{\rm coll}$ determination in peripheral collisions.
Semi-inclusive h+jet $I_{\rm CP}$ measurements are also below unity.
However, it remains to be studied whether effects beyond jet quenching contribute to this suppression, such as nPDF effect, $\alpha$-cluster structure effect, or EA-$Q^2$ anti-correlation.
Future studies could include similar measurements in $p$+$p$ collisions to provide a reference for the calculation of jet $R_{\rm AA}$, and semi-inclusive analyses could compare trigger-side and recoil-side to investigate potential anti-correlations.
Comparison with theoretical models and upcoming LHC O+O results will also be valuable to further understanding.

\section{Acknowledgments}

The speaker is supported by the National Natural Science Foundation of China (NSFC) under Grant Nos. 12475144, 11890710 and 11890713.

%
\bibliography{references} 
%

\end{document}